\documentclass[authoryear,preprint,review,11pt]{elsarticle}

\usepackage{amssymb}
\usepackage{amsmath}
\usepackage{mathtools}
\usepackage{IEEEtrantools}
\usepackage[table,xcdraw,svgnames,dvipsnames]{xcolor}
\usepackage{url,lineno,microtype,subcaption}
\usepackage[onehalfspacing]{setspace}
\usepackage{bm}
\usepackage{scalerel}
\usepackage{mathrsfs}
\usepackage{relsize}
\usepackage{textcomp}
\usepackage{gensymb}
\usepackage{graphicx}
\usepackage[colorlinks]{hyperref}
\usepackage{upgreek}
\usepackage{textgreek}
\usepackage[T1]{fontenc}
\usepackage{parskip}
\usepackage{media9}
\usepackage{listings}
\usepackage{float}
\usepackage{algorithm}
\usepackage[noend]{algpseudocode}
\usepackage{textcomp}
\usepackage{comment}

\definecolor{cerulean}{rgb}{0.0, 0.48, 0.65}
\AtBeginDocument{%
  \hypersetup{
    citecolor=cerulean,
    linkcolor=Red,   
    urlcolor=Magenta}}
    
\graphicspath{ {./figures/} }

\definecolor{mygreen}{RGB}{28,172,0} 
\definecolor{mylilas}{RGB}{170,55,241}
\journal{\,}

\begin{document}
\begin{frontmatter}

\title{Evaluating phase synchronization methods in fMRI: a comparison study and new approaches}


\author[A1]{Hamed Honari}
\author[A2,A3,A4]{Ann S. Choe}
\author[A5]{Martin A. Lindquist\corref{cor1}}
\cortext[cor1]{Corresponding author}
\ead{mlindqui@jhsph.edu}
\address[A1]{Department of Electrical and Computer Engineering, Johns Hopkins University, USA}
\address[A2] {F. M. Kirby Research Center for Functional Brain Imaging, Kennedy Krieger Institute, USA}
\address[A3] {International Center for Spinal Cord Injury, Kennedy Krieger Institute, USA}
\address[A4] {Russell H. Morgan Department of Radiology and Radiological Science, Johns Hopkins School of Medicine, USA}
\address[A5]{Department of Biostatistics, Johns Hopkins University, USA}

\begin{abstract}
In recent years there has been growing interest in measuring time-varying functional connectivity between different brain regions using resting-state functional magnetic resonance imaging (rs-fMRI) data. One way to assess the relationship between signals from different brain regions is to measure their phase synchronization (PS) across time.  There are several ways to perform such analyses, and here we compare methods that utilize a PS metric together with a sliding window, referred to here as windowed phase synchronization (WPS), with those that directly measure the instantaneous phase synchronization (IPS). In particular, IPS has recently gained popularity as it offers single time-point resolution of time-resolved fMRI connectivity. In this paper, we discuss the underlying assumptions required for performing PS analyses and emphasize the necessity of band-pass filtering the data to obtain  valid results. We review various methods for evaluating PS and introduce a new approach within the IPS framework denoted the cosine of the relative phase (CRP). We contrast methods through a series of simulations and application to rs-fMRI data. Our results indicate that CRP outperforms other tested methods and overcomes issues related to undetected temporal transitions from positive to negative associations common in IPS analysis. Further, in contrast to phase coherence, CRP unfolds the distribution of PS measures, which benefits subsequent clustering of PS matrices into recurring brain states.

\end{abstract}


\end{frontmatter}




\section{Introduction}\label{Intro}

It was previously assumed that functional connectivity (FC) in the brain was static during the course of a single resting-state functional magnetic resonance imaging (rs-fMRI) run. Recently, however, several studies \citep{chang2010time, hutchison2013dynamic, preti2016dynamic, tagliazucchi2012dynamic, thompson2013short, allen2014tracking, lurie2019questions} have pointed to dynamic changes in FC taking place in a considerably shorter time window than  previously thought (i.e., on the order of seconds and minutes). Several methods have been proposed to investigate such time-varying connectivity (TVC). These include the widely-used sliding-window approach \citep{tagliazucchi2012dynamic, chang2010time, hutchison2013dynamic, hutchison2013resting}, change point analysis  \citep{cribben2012dynamic, cribben2013detecting, xu2015dynamic}, point process analysis \citep{tagliazucchi2011spontaneous}, co-activation patterns (CAPs) \citep{liu2013time}, transient-based CAPs \citep{karahanouglu2015transient}, time series models \citep{lindquist2014evaluating}, time-frequency analysis \citep{chang2010time}, and variants of hidden Markov models (HMMs)  \citep{eavani2013unsupervised, vidaurre2017brain, shappell2019improved, bolton2018interactions}.  Despite development of these promising approaches, estimation of TVC remains a challenging endeavor due to the low signal-to-noise ratio (SNR) of the blood oxygen level dependent (BOLD) signal and the presence of image artifacts and nuisance confounds \citep{hutchison2013dynamic, lindquist2014evaluating, laumann2016stability}.

The term {\em synchronization} refers to the coordination in the state of two or more systems that can be attributed to their interaction (or coupling) \citep{rosenblum1996phase}. Recently, phase synchronization (PS) methods were proposed as a means of measuring the level of synchrony between time series from different regions of interest (ROIs) in the brain \citep{glerean2012functional, pedersen2017spontaneous, pedersen2018relationship}.  Typically, the phase of a particular time series is computed at each time point through the application of the Hilbert transform, and used to evaluate the phase difference between various time series. Two time series in synchronization will maintain a constant phase difference.
In this study, we differentiate between methods that combine a PS metric with a sliding window approach, referred to as windowed phase synchronization (WPS), with those that directly measure PS at each time point, referred to as instantaneous phase synchronization (IPS). 

The first class of methods (WPS methods) uses metrics that provide a single omnibus measure of the phase synchronization between two time series obtained using Hilbert Transform. This approach is similar to how correlations provide an omnibus measure of the linear relationship between time series (analogous to the static correlation used in FC). In this approach, a sliding window technique is used to compute the metric locally within a specific time window. As the window is shifted across time, one can obtain a time-varying value of the measure of interest (i.e., the dynamic synchronization between two time series). The use of Phase Locking Value (PLV) \citep{glerean2012functional, boccaletti2018synchronization, pauen2013circular} to capture time-varying relationship between a pair of signals has recently been used in this context \citep{rebollo2018stomach}.  In this paper, we propose two other measures that can capture the time-varying relationship between a pair of signals:  circular-circular correlation \citep{jammalamadaka2001topics, pauen2013circular}, and toroidal-circular correlation \citep{zhan2017circular}. Importantly, this class of methods suffers from similar issues as sliding-window correlations, such as the need to select an \textit{a priori} window length for analysis. 

The second class of methods (IPS methods) directly analyzes the instantaneous phases extracted using the Hilbert Transform.  In recent years there has been growing interest in using IPS methods in neuroimaging, with the bulk of the work applied to MEG and EEG data.  However, several studies have also applied IPS methods to fMRI data.  For instance, \cite{laird2002characterizing} used IPS methods to analyze task-activated fMRI data. However, the lack of narrow band-pass filtering in the study's analysis pipeline brings into question the validity of the results.  \cite{niazy2011spectral} studied the spectral characteristics of resting state network (RSN) and suggested that it is important to consider the IPS between various RSNs at different frequencies. \cite{glerean2012functional} proposed using IPS as a measure of TVC. 

Finally, \cite{pedersen2018relationship} examined the relationship between IPS and Correlation-based Sliding Window (CSW) techniques and observed a strong association between the two methods when using absolute values of CSW.
Benefits of using an IPS approach is that it offers single time-point resolution of time-resolved fMRI connectivity, and does not require choosing an arbitrary window length.

In this paper, we discuss the concept of phase synchronization in the context of fMRI, with a particular focus on TVC. We begin by reviewing the framework for computing the phase from time series data using the Hilbert transform, and discuss the necessity of band-pass filtering the data to accurately estimate the instantaneous phases. We continue by introducing a number of different methods for evaluating phase synchronization. We focus both on methods already in common use, such as the phase locking value and phase coherence, as well as methods new to the fMRI literature, such as circular-circular correlation and toroidal-circular correlation \citet{zhan2017circular}. Finally, we propose a new variant of phase coherence, denoted the Cosine of the Relative Phase (CRP), that can be used to compute the IPS. We contrast these methods through a series of simulations and application to rs-fMRI data.

\section{Methods}

\subsection{A Framework for Computing Instantaneous Phase}\label{sec: concept of phase sync}

To obtain the instantaneous phase \citep{boccaletti2018synchronization} of an arbitrary real signal $x(t)$ one must first construct an analytic signal:
\begin{equation}
    z(t) = x(t) + j \mathcal{H}\{x(t)\}
\end{equation}
where $j = \sqrt{-1}$ and $\mathcal{H}$ represents the Hilbert Transform.   
This signal can in turn be re-expressed as follows:
\begin{equation}
    z(t) = \mathfrak{A}(t) \exp\big(j\mathfrak{\phi}(t)\big)
\end{equation}
where $\mathfrak{A}(t)$ represents the envelope and $\mathfrak{\phi}(t)$ the instantaneous phase.
 
Here $x(t)$ is assumed to satisfy Bedrosian's Product Theorem, which states that a band-limited signal can be decomposed into the product of envelope and phase when their spectra are disjoint.  This holds if the signal of interest is first narrow-banded by applying a band-pass filter.

There are two important considerations when choosing the appropriate filter to apply.  First, it should not corrupt the phase information in the signal. Thus, it is important to use a filter that does not shift the phase. One class of filters that accomplishes this goal are zero-phase filters.  Second, the width of the frequency band must be sufficiently narrow.  

The narrower the band, the closer the signal will be to a monocomponent signal and the Hilbert transform will produce an analytic signal with meaningful envelope and phase.  

The choice of appropriate band widths in this context have been investigated in previous studies of fMRI data. For example, \citet{ponce2015resting} examined band-pass filtering of fMRI data using various frequency bands in the range of $0.01-0.13 \, Hz$, and reported consistent results for phase statistics at each frequency band. \citet{pedersen2018relationship} compared using a narrow band-pass filter ($0.03-0.07 \, Hz$) with a wider band-pass filter ($0.01-0.1 \, Hz$), and found that the narrow-band data yielded stronger associations between the results of CSW and IPS analyses.

\begin{figure}[!ht]
\begin{center}
\includegraphics[width=0.8\textwidth]{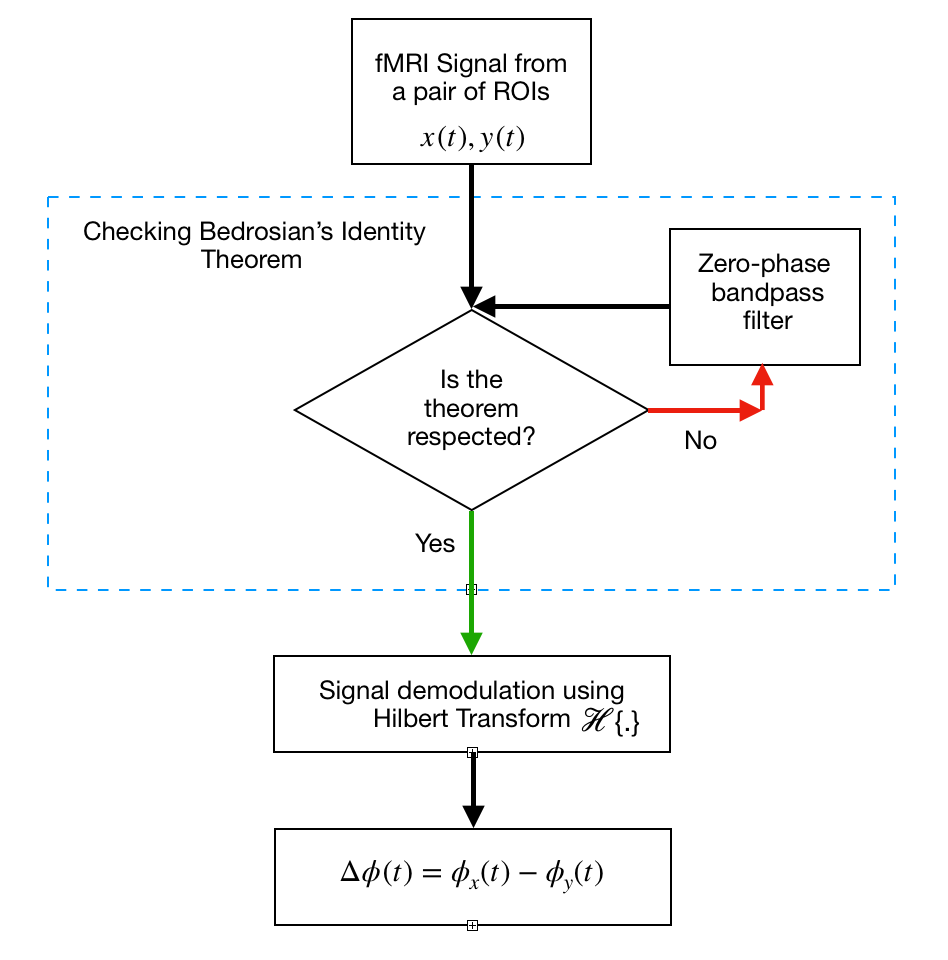}
\end{center}
\caption{A schematic of the approach to calculate the  instantaneous phase (IP) framework}\label{fig: fig1}
\end{figure}

A schematic framework for obtaining the instantaneous phase synchronization is shown in Figure \ref{fig: fig1}.
Consider that a pair of time series $x(t)$ and $y(t)$, $t=1, \ldots T$, from two different ROIs are filtered using a narrow band-pass zero-phase filter, $h_{bp}(t)$, and denote the filtered data by $x_n(t)$ and $y_n(t)$ respectively, i.e.,
    \begin{align}
           x_n(t) &= x(t)*h_{bp}(t)\\
           y_n(t) &= y(t)*h_{bp}(t) .
    \end{align}
Here $*$ represents the convolution operator.

If Bedrosian's theorem holds, the analytical signals of the narrow-banded time series can be expressed as the product of instantaneous envelope and instantaneous phase:  
\begin{align}
    x_a(t) = x_n(t) + j \mathcal{H}\{x_n(t)\} =\mathfrak{A}_{x_n}(t)\exp\big(j\mathfrak{\phi}_x(t)\big)\\
    y_a(t) = y_n(t) + j \mathcal{H}\{y_n(t)\} =\mathfrak{A}_{y_n}(t)\exp\big(j\mathfrak{\phi}_y(t)\big).
\end{align}
Here the subscript $a$ refers to analytical signal. Throughout, we assume that $\phi_x(t)$ and $\phi_y(t)$ are the phase time series extracted from a pair of time series $x(t)$, and $y(t)$.  Using the instantaneous phases, synchronization can be assessed by studying their differences.

\subsection{Methods for Assessing Phase Synchronization}\label{sec: PSAssesmentMeasures}

Next, we describe how to measure PS based on the extracted phase time series.  We discriminate between methods that utilize a PS metric together with a sliding window approach (WPS) from those that directly measure IPS.  

\subsubsection{Windowed Phase Synchronization}

The first class of methods, place a measure of PS across two time series within a sliding window framework. Here we describe this approach using PLV, circular-circular correlation, and toroidal circular correlation.

\paragraph{Phase Locking Value}\hspace{1cm}

The PLV is a classic metric for assessing phase synchronization based on quantifying to what extent the two signals are phase locked. PLV has found widespread use in the analysis of MEG/EEG data \citep{rosenblum2000detection, halliday1998using}.  

This notion of synchronization can be expressed as follows:
\begin{equation}\label{locking condition}
    |\Delta\Phi_{m,n}(t)| < const, \;\; \textrm{where}\; \Delta\Phi_{m,n}(t) = n \phi_x(t) - m\phi_y(t)  .
\end{equation}
Here the integers $m$ and $n$ are the synchronization indices and $\Delta\Phi_{m,n}(t)$ the generalized phase difference time series. 

In this paper, we assume $m = n = 1$ and drop the indices and let
$\Delta\Phi_{m,n}(t) = \Delta \Phi(t)$.

Using the instantaneous phase difference of the signals at each time point, the PLV can be computed as follows:
\begin{equation}\label{eqn: PLV}
    PLV = \bigg| \Big< e^{j\mathbf{\Delta\mathbf{\phi}(t)}}  \Big>_t \bigg|
\end{equation}
\noindent where the operator $\big<\cdot\big>_t$ denotes averaging over time.  If the pair of signals are unsynchronized, then $PLV=0$ and $\mathbf{\Delta\phi(t)}$ follows a uniform distribution; otherwise, if the pair are synchronized, $PLV$ is constant and equal to $1$ \citep{lachaux1999measuring}. 

To compute PLV within a sliding window framework, for each time window of length $\ell$, the PLV between the pair of the signals can be obtained using Equation (\ref{eqn: PLV}).  This approach has been previously used for assessing the episodes of elevated gastric-BOLD synchronization by \citep{rebollo2018stomach} in the study of stomach-brain synchrony.  

Next, we introduce two other measures in WPS approach that to our best of knowledge have not been used to assessing the time varying phase synchornization in a sliding window fashion.

\paragraph{Circular-Circular Correlation}\label{circcircStatdef}\hspace{1cm}

The instantaneous phase obtained from each time series are directional data and follow a circular distribution.  In this context, the use of the standard Pearson correlation coefficient is no longer appropriate.  Instead, a more suitable measure is circular-circular correlation \citep{jammalamadaka1988correlation, jammalamadaka2001topics}, defined as follows:
\begin{equation}\label{eqn:CircCorr}
    \rho_{circ}= \frac{\mathbb{E}\Big[sin( {\Phi}_x-\mu)sin(\Phi_y-\nu)\Big]}{\sqrt{\mathbb{E}\Big[sin^2(\Phi_x-\mu)\Big]  \mathbb{E}\Big[sin^2(\Phi_y-\nu)\Big]}}.
\end{equation}
\noindent In the equation above, $\Phi_x = (\phi_x(1), \ldots \phi_x(T))$ and $\Phi_y = (\phi_y(1), \ldots \phi_y(T))$, while $\mu$ and $\nu$ represent the mean directions of $\Phi_x$ and $\Phi_y$, respectively.  Thus, the terms $sin(\Phi_x-\mu)$ and $sin(\Phi_y-\nu)$ can be interpreted as the deviations of $\Phi_x$ and $\Phi_y$ from their corresponding mean directions.

The circular-circular correlation provides a measure of the static interdependence between the two phase time series. It can also be used within the sliding windows framework to investigate the time-varying PS.  This can be expressed as:    
\begin{equation}
    \hat{\rho}_{circ,t} =  \frac{\mathlarger{\sum}_{s=t-\ell-1}^{t-1}\Big[sin(\phi_{x}(s)-\hat{\mu}_t)sin(\phi_{y}(s)-\hat{\nu}_t)\Big] }{\sqrt{\Big(\mathlarger{\sum}_{s=t-\ell-1}^{t-1}\Big[sin^2(\phi_{x}(s)-\hat{\mu}_t)\Big] \Big)
    \Big(\mathlarger{\sum}_{s=t-\ell-1}^{t-1}
    \Big[sin^2(\phi_{y}(s)-\hat{\nu}_t)\Big]\Big)}},
    \label{circcoeffestimate}
\end{equation}
where $\hat{\mu}_t$ and $\hat{\nu}_t$ represent the estimated time-varying mean of the two phase time series over the sliding window.  

It is important to note in the context of directional statistics, 

$\hat{\mu}_t$ is computed as follows \citep{jammalamadaka2001topics, bishop2006pattern}:
\begin{equation} \label{ccmean}
\hat{\mu}_t = \tan^{-1}\Bigg\{\frac{\mathlarger{\sum}_{s=t-\ell-1}^{t-1} \sin{\phi_{x}(s)}}{\mathlarger{\sum\limits_{s=t-\ell-1}^{t-1}\cos{\phi_{x}(s)}}} \Bigg\}
\end{equation}
\noindent  This formulation can be understood by representing each directional variable on a unit circle ($\mathrm{r}=1$) in the polar coordinate system $(\mathrm{r},\phi)$. Re-expressing to the Cartesian coordinate system $(\mathrm{x},\mathrm{y})$, we can write $\cos\phi_x(i) = \mathrm{x}_i$ and $\sin\phi_y(i) = \mathrm{y}_i$, for $i = 1,...,n$. Thus, $\bar{\mathrm{x}} = n^{-1}\sum{\cos\phi_x(i)}$, $\bar{\mathrm{y}} = n^{-1} \sum{\sin\phi_y(i)}$, and the mean direction thus can be written as expressed in Eq. \ref{ccmean}.

Here $\hat{\nu}_t$ is computed analogously.

\paragraph{Toroidal circular correlation}\label{sec: newCircCorr}\hspace{1cm}

In a critique of circular-circular correlation,
\citet{zhan2017circular} argued that the sine of an angle contains less information than the angle itself, as multiple angles can take the same sine value. Furthermore, since the sine function is not monotone within an interval of $\pi$, this may lead to unreasonable results.  To circumvent these issues, they introduced a circular correlation coefficient for bivariate directional data on a torus, which is the equivalent to the product of two circles \citep{sojakova2016equivalence, zhan2017circular}.

To elaborate, let $\phi_{x}(t_1)$ (or similarly $\phi_{y}(t_1)$) and $\phi_{x}(t_2)$ (or $\phi_{y}(t_2)$) be two circular data points and set $0 \leq \phi_{x}(t_1),\phi_{x}(t_2)<2\pi$, so $|\phi_{x}(t_1) - \phi_{x}(t_2)|<2\pi$.
When $-\pi<\phi_{x}(t_1)-\phi_{x}(t_2)\leq 0$ OR $\phi_{x}(t_1)-\phi_{x}(t_2) > \pi$, by convention the direction from $\phi_{x}(t_1)$ to $\phi_{x}(t_2)$ is considered to be clockwise. When $\phi_{x}(t_1)-\phi_{x}(t_2)\leq -\pi$ OR $0<\phi_{x}(t_1)-\phi_{x}(t_2)\leq\pi$, the direction is considered counter-clockwise. The same definition holds for $\phi_{y}(t_1)$ and $\phi_{y}(t_2)$.\vspace{\baselineskip}

Let $\delta = \phi_{x}(t_1)-\phi_{x}(t_2)$, then the order function can be expressed as follows:
\begin{equation}\label{eqn: ordfunc2}
    h(\delta) = \Big[(\delta+2\pi) \;\;\;\; mod\;\: 2\pi \Big] - \pi =     \begin{cases}
    \begin{aligned}
        &\delta + \pi, \qquad  &-2\pi<\delta<0,\\
        \\
        &\delta-\pi, \qquad &0\leq \delta<2\pi.
    \end{aligned}
    \end{cases}
\end{equation}

Now, let us assume that $(\phi_{x}(t_1), \phi_{y}(t_1))$ and $(\phi_{x}(t_2),\phi_{y}(t_2))$ are independent. The circular correlation is then defined as follows:
\begin{equation}
    \rho_{tor} = \frac{   \mathbb{E} \Big[ h(\phi_{x}(t_1),\phi_{x}(t_2)) h(\phi_{y}(t_1),\phi_{y}(t_2))\Big] } {\sqrt{\mathbb{E}\Big[h(\phi_{x}(t_1),\phi_{x}(t_2))^2\Big]  \mathbb{E}\Big[h(\phi_{y}(t_1),\phi_{y}(t_2))^2\Big]}}
\end{equation}

Based on this definition, the two variables $\phi_x$ and $\phi_y$ move on the circumference of the torus in the same direction if $h(\phi_{x}(t_1),\phi_{x}(t_2)) h(\phi_{y}(t_1),\phi_{y}(t_2))>0$, making $\rho_{tor}>0$.  Similarly, $\rho_{tor}<0$ indicates that the two variables are moving in opposite directions.

An estimator can be obtained as follows:
\begin{equation}\label{toroidalcoeffestimate}
    \hat{\rho}_{tor} = \frac{\mathlarger{\sum}_{1\leq t_i\leq t_j \leq n}\Big[h(\phi_{x}(t_i),\phi_{x}(t_j)) h(\phi_{y}(t_i),\phi_{y}(t_j))\Big]}{\sqrt{\Big(\mathlarger{\sum}_{1\leq t_i\leq t_j \leq n}h(\phi_{x}(t_i),\phi_{x}(t_j))^2\Big) \Big( \mathlarger{\sum}_{1\leq t_i\leq t_j \leq n}h(\phi_{y}(t_i),\phi_{y}(t_j))^2\Big)}}
\end{equation}

The main advantage of using toroidal circular correlation is that no information about the angles are lost.  Thus, the estimator circumvents problems due to the non-monotonicity of the sine function that could result in irregular estimation when using the circular-circular correlation.  It can be calculated within a sliding window framework similar to the previous sections.

\subsubsection{Instantaneous Phase Synchronization}

The second class of methods are based on directly working with the 
instantaneous phases of two time series. Here we focus on phase coherence \citep{pedersen2017spontaneous, pedersen2018relationship}, which has already found wide usage in the field, and the cosine of the relative phase, introduced for the first time in this work.

\paragraph{Phase Coherence}\hspace{1cm}

The phase coherence at each time point is defined as follows: \begin{equation}\label{eqn: phase coherence}
    \Psi(t) = 1 - \Big|\sin{(\Delta\Phi(t))}\Big|
\end{equation}
Here the absolute value of the sine of the relative phase differences is included to account for phase wrapping and resolve issues with phase ambiguity over time.   Note that the range of the values obtained using this metric will take values between $0$ and $1$, where $0$ implies no phase coherence and $1$ corresponds to maximal phase coherence.  

A shortcoming of this approach is that it discards information about the direction of the relationship as $|\sin(-\Delta\Phi)| = |\sin(\Delta \Phi)|$. As these values vary between 0 and 1, $\Psi(t)$ as defined in Eq. \ref{eqn: phase coherence} does not capture negative association (i.e., when signals are in anti-phase). This may help explain why \citet{pedersen2018relationship} found that the association between IPS and CSW analysis was strongly dependent on negative correlations obtained from the CSW analysis, and that the association increased when comparing the absolute values of the correlations.  In addition, it explains why their analysis was unable to capture temporal transitions from positive to negative associations, and vice versa, that appeared in the CSW analysis.  

\paragraph{Cosine of the Relative Phase}\hspace{1cm}

To circumvent the issues outlined above, we propose a modification of phase coherence that takes temporal transitions into account and preserves the correlational structure in the data.  This can be achieved by not taking the absolute value of the phase difference and using a cosine function instead of a sine function. We refer to this measure as the cosine of the relative phase (CRP), defined as follows:
\begin{equation}
    \vartheta(t) = \cos{(\Delta\Phi(t))}
\end{equation}
Notably, the range of the values obtained using this metric take values between $-1$ and $1$, and is therefore directly comparable to standard correlation values. 

The CRP approach avoids phase unwrapping and takes phase ambiguity into consideration.  When the instantaneous phase of two signals are similar to one another (i.e., $|\Delta\Phi(t)|\approx 0$), CRP yields a value close to 1.  When the phases are dissimilar but in the same direction, their relative phase difference is bounded between [$-\pi/2$,$\pi/2$], which is the range where the cosine function is positive.  As the phases become orthogonal to one another, CRP approaches $0$ indicating a lack of coherence.  Similarly, the CRP captures negative associations between phases.  If the phase difference is greater than $\pm \pi/2$, this results in negative values of the cosine function.  Thus, using CRP as a measure of phase synchrony helps overcome the issue of detecting temporal transitions from positive to negative associations (or vice-versa), and preserves the positive and negative dependence in the data.

\subsection{Simulations}\label{sec: simulation}

In this section we introduce three simulations designed to compare the methods presented in Section \ref{sec: PSAssesmentMeasures} for the two different classes of PS analysis.  The first investigates their performance in a null setting, while the second and third investigate PS measures when two sinusoidal signals have the same frequency but differing phase shifts. For all three simulations data was generated with a sampling frequency of $1/TR$, where $TR$ represents the repetition time of an fMRI experiment. To be comparable with the rs-fMRI data used in this paper we chose $TR=2$ seconds. 

For each simulation we computed WPS values using the PLV, circular-circular correlation, and toroidal correlation, and IPS values using phase coherence and the CRP method.  All simulations were repeated $1000$ times, and the mean and variance of the PS measured at each time point was used to construct a $95\%$ confidence interval.  Furthermore, the effect of different window lengths in the WPS analysis was evaluated using three different window lengths ($30$, $60$, and $120$ TRs). 

To illustrate the necessity of band-pass filtering the data, PS analysis was performed on the simulated data both before and after band-pass filtering it in the range $[0.03,  0.07] \, Hz$.  
Throughout we used a $5^{th}$ order Butterworth filter.  The zero-phase version of this filter is implemented in MATLAB by filtering backward in time using MATLAB's \verb+filtfilt+ function to cancel out the phase delay introduced by this filter.

\underline{\em Simulation 1:}  \,\, To simulate time series with independent phase dynamics, we generated two independent random signals from a Gaussian distribution with mean $0$ and standard deviation $1$. 

Using the logic of surrogate data testing, we generated surrogate data under the assumption of no relationship between the phase from the two signals. To achieve this goal we used cyclic phase permutation (CPP) surrogates \citep{lancaster2018surrogate}, constructed by reorganizing the cycles within the extracted phase of the signals. This destroys any phase dependence between the pair, whilst preserving the general form of the phase dynamics of each time series. For this simulation, the $1000$ realizations of signal pairs were generated using CPP surrogates.

\underline{\em Simulation 2:} \,\, Here we generated two sinusoidal signals with the same frequency, but with a time-varying phase shift corresponding to a ramp function. To elaborate, consider two sinusoidal signals $x(t)$ and $y(t)$. Let $x(t)$ be the reference signal with an angular frequency of $\omega_0$ and phase $\varphi_x(t)$. Further, let $y(t)$ have the same angular frequency but with phase $\varphi_y(t)$. The signals can be expressed as follows:
\begin{equation}\label{eqn: signal phase shift}
    \begin{split}
        x(t) &= A_x cos\big(\omega_0 t + \varphi_x(t)\big) + \varepsilon_{x}\\
        y(t) &= A_y cos\big(\omega_0 t + \varphi_y(t) \big) + \varepsilon_{y}
    \end{split}
\end{equation}

Without loss of generality, let $\varphi_x(t)=0$ and $\varphi_y(t)$ be a ramp function, 
\begin{equation}\label{eq: rampfunction}
    r(t-t_0) = 
    \begin{cases}
        \begin{aligned}
            &0 \qquad  &t &\leqslant t_0\\
            &t-t_0 \qquad &t &> t_0 
            \end{aligned}
        \end{cases}
\end{equation}
The time series can then be expressed as follows:
\begin{equation}
    \begin{split}
        x(t) &= A_x cos(\omega_0 t) + \varepsilon_{x}\\
        y(t) &= A_y cos\Bigg(\omega_0 t + 4\pi r(t-t_0) \Bigg) + \varepsilon_{y}
\end{split}
\end{equation}
Throughout, $\omega_0 = 2\pi f \;rad/s$ with $f=0.05 Hz$, and the transition is set to occur at $t_0 = 170 \, s$.  The noise terms $\varepsilon_{x}$ and $\varepsilon_{y}$ are Gaussian white noise with mean $0$ and standard deviation $1$.  

To summarize, the two signals start out phase synchronized and remain in this state up to $t_0 = 170 \, s$.  After which the phase  difference starts linearly increasing and transition into a non-synchronized state.

{\em Simulation 3:} \, \, Here we generated two sinusoidal signals with the same frequency, but with a time-varying phase shift corresponding to a sigmoid function. As in the previous simulation, data was generated according to Eq. (\ref{eqn: signal phase shift}). Here we let $\varphi_x(t)=0$ and $\varphi_y(t)$ be a sigmoid function, i.e. 
\begin{equation}
    s(t-t_0) = \frac{a}{1+\exp{\big(b(t-t_0)\big)}}
\end{equation}
Hence, the time series can be expressed as follows:
\begin{equation}
    \begin{split}
        x(t) &= A_x cos(\omega_0 t) + \varepsilon_{x}\\
        y(t) &= A_y cos\Bigg(\omega_0 t + \frac{a}{1+\exp{\big(b(t-t_0)\big)}} \Bigg) + \varepsilon_{y}
\end{split}
\end{equation}
Throughout, we set $a=2\pi$, $b=-0.01$, $t_0 =170$, and $\omega_0 = 2\pi f \;rad/s$ with $f=0.05 Hz$. The noise terms $\varepsilon_{x}$ and $\varepsilon_{y}$ are Gaussian white noise with mean $0$ and standard deviation $1$.  

To summarize, the signals are initially in phase, after which the amount of phase shift gradually increases. This continues until $t=170$ when the pairs are in anti-phase synchronization. Thereafter, the signals gradually return to being in phase.  The transition between the phase of the signals from 0 to $2\pi$ occurs smoothly and monotonically increasing.

\subsection{Application to Kirby data set}

\subsubsection{Image Acquisition}

We used the Multi-Modal MRI Reproducibility Resource from the F.M. Kirby Research Center, commonly referred to as Kirby21. It includes data from $21$ healthy adults scanned on a 3T Philips Achieva scanner designed to achieve $80$ $mT/m$ maximum gradient strength with body coil excitation and an eight channel phased array SENSitivity Encoding (SENSE) \citep{pruessmann1999sense} head-coil for reception. All subjects completed two scanning sessions on the same day, between which they briefly exited the scan room. A T1-weighted (T1w) Magnetization-Prepared Rapid Acquisition Gradient Echo (MPRAGE) structural run was acquired during both sessions (acquisition time = $6$ $min$, TR/TE/TI = 6.7/3.1/842  $ms$, resolution = $1\times1\times1.2 \, mm^3$, SENSE factor = $2$, flip angle = $8\degree$). A multi-slice SENSE-EPI pulse sequence \citep{stehling1991echo, pruessmann1999sense} was used to acquire two rs-fMRI runs during each session. Each run consisted of 210 volumes sampled every $2 \, s$  at $3 \, mm$ isotropic spatial resolution (acquisition time: $7 \, min$, TE = 30 $ms$, SENSE acceleration factor $= 2$, flip angle = $75\degree$, $37$ axial slices collected sequentially with a $1 \, mm$ gap). Subjects were instructed to rest comfortably while remaining still. One subject was excluded from further analyses due to excessive head motion. For a more detailed description of the acquisition protocol see \citet{landman2011multi}.

\subsection{Image Processing}

The data was preprocessed using SPM8 (Wellcome Trust Centre for Neuroimaging, London, United Kingdom) \citep{friston1994statistical} and custom MATLAB (The Mathworks, Inc., Natick, MA) scripts. Five initial volumes were discarded to allow for the stabilization of magnetization. Slice-time correction was performed using as a reference the slice acquired at the middle of the TR. Rigid body realignment transformation was performed to adjust for head motion. Structural runs were registered to the first functional frame and normalized to Montreal Neurological Institute (MNI) space using SPM8's unified segmentation-normalization algorithm \citep{ashburner2005unified}. The estimated nonlinear spatial transformations were applied to the rs-fMRI data, which were high-pass filtered using a cutoff frequency of $0.01 \, Hz$. The rs-fMRI data was spatially smoothed using a $6 \, mm$ full-width-at-half-maximum (FWHM) Gaussian kernel, which is twice the nominal size of the rs-fMRI acquisition voxel. 

The Group ICA of fMRI toolbox (GIFT) (\url{https://trendscenter.org/software/gift/}) was used to estimate the number of independent components (ICs) present in the data, perform data reduction via principal component analysis (PCA) prior to independent component analysis (ICA), and perform group independent component analysis (GICA) \citep{calhoun2001method} on the PCA-reduced data. The number of ICs was estimated using the minimum description length (MDL) criterion \citep{li2007estimating}. Across subjects and sessions, 56 was the maximum estimated number of ICs and 39 the median. Prior to GICA, the image mean was removed from each time point for each session, and three steps of PCA were performed. Individual session data were reduced to 112 principal components (PCs), which were concatenated within subjects in the temporal direction and further reduced to 56 PCs. Finally, data were concatenated across subjects and reduced to 39 PCs. The dimensions of the individual session PCA (112) was chosen by doubling the estimated maximum IC number (56), to ensure robust back-reconstruction \citep{allen2011baseline, allen2012capturing} of subject- and session-specific spatial maps and time courses from the group-level ICs. ICA was repeated on these 39 group-level principal components 10 times, utilizing the Infomax algorithm with random initial conditions \citep{bell1995information}. The resulting 390 ICs were clustered across iterations using a group average-link hierarchical strategy, and 39 aggregate spatial maps were defined as the cluster modes. Subject- and session-specific spatial maps and time courses were generated from these aggregate ICs using the GICA3 algorithm.  

The spatial distribution of each IC was compared to a publicly available set of 100 unthresholded t-maps of ICs estimated using rs-fMRI data collected from 405 healthy participants \citep{allen2014tracking}. These maps were pre-classified as resting-state networks (RSNs) or noise by a group of experts, and the 50 components classified as RSNs have been organized into seven functional groups, namely visual (Vis), auditory (Aud), somatomotor (SM), default mode (DMN), cognitive-control (CC), sub-cortical (SC) and cerebellar (Cb) networks. For each spatial map, we calculated the percent variance explained by each of the seven sets of RSNs. The functional assignment of each Kirby component was determined by the set of  components that explained the most variance, and if the top two sets of RSNs explained less than 50\% of the variance in a Kirby component, the component was labeled as noise. In total 21 of the 39 components were assigned to a RSN. Subject- and run-specific time series from these components served as input for our analyses. 

\subsubsection{Analysis}\label{sec: analysisKirby21}

Data from a single run of the Kirby21 was used.  Thus, the data consisted of $21$ ROIs measured over $210$ time points for $20$ subjects.  The framework described in Section \ref{sec: concept of phase sync} (see Fig. \ref{fig: fig1}), using a band-pass filter with range $[0.03,  0.07] \, Hz$, was applied to the data to compute the region-wise instantaneous phase for each of the $20$ subjects. For each pair of subject-specific phase time series, we applied the WPS and IPS methods. We also applied CSW for comparison purposes. To facilitate comparison between the WPS and CSW methods, we used a common window length of $28$ time points. We further compared the results with a prewhitened Correlation-based Sliding Window (PW-CSW) assuming an AR(1) model.  This comparison was performed as a previous study  \citep{honari2019investigating} showed that prewhitening the data prior to analysis can lower the variance of the estimated TVC and improve brain state estimation.

Application of each method gave rise to a subject-specific $21 \times 21 \times 210$ array of PS measures. Following the approach of Allen and colleagues (\cite{allen2014tracking}), we applied $k$-means clustering to estimate recurring brain states across subjects.  First, we reorganized the lower triangular portion of each subject's dynamic correlation data into a matrix of dimension $210 \times 210$. Here the row dimension corresponds to the number of elements in the lower triangular portion of the matrix (i.e., $21(21-1)/2$), and the column dimension corresponds to the number of time points. Then we concatenated the data from all subjects into a matrix with row dimensions $210$ and column dimensions $(210 \times 20 = 4200)$. Finally, we applied $k$-means clustering to the concatenated data, where each of the resulting cluster centroids were assumed to represent a recurring brain state.  The k-means clustering was repeated 200 times, using random initialization of centroid positions, in order to increase the chance of escaping local minima.  In this study, we set the number of centroids to two, representing two distinct brain states, as determined using the Davies-Bouldin Index (DBI) \citep{davies1979cluster}.  This is consistent with the number of the clusters for this dataset used in previous studies by \cite{choe2017comparing} and \cite{honari2019investigating}.  

\section{Results}

\subsection{Simulation 1}

\begin{figure}[htb]
\begin{center}
\includegraphics[width=\textwidth,trim={2.5cm 0cm 1cm 0},clip]{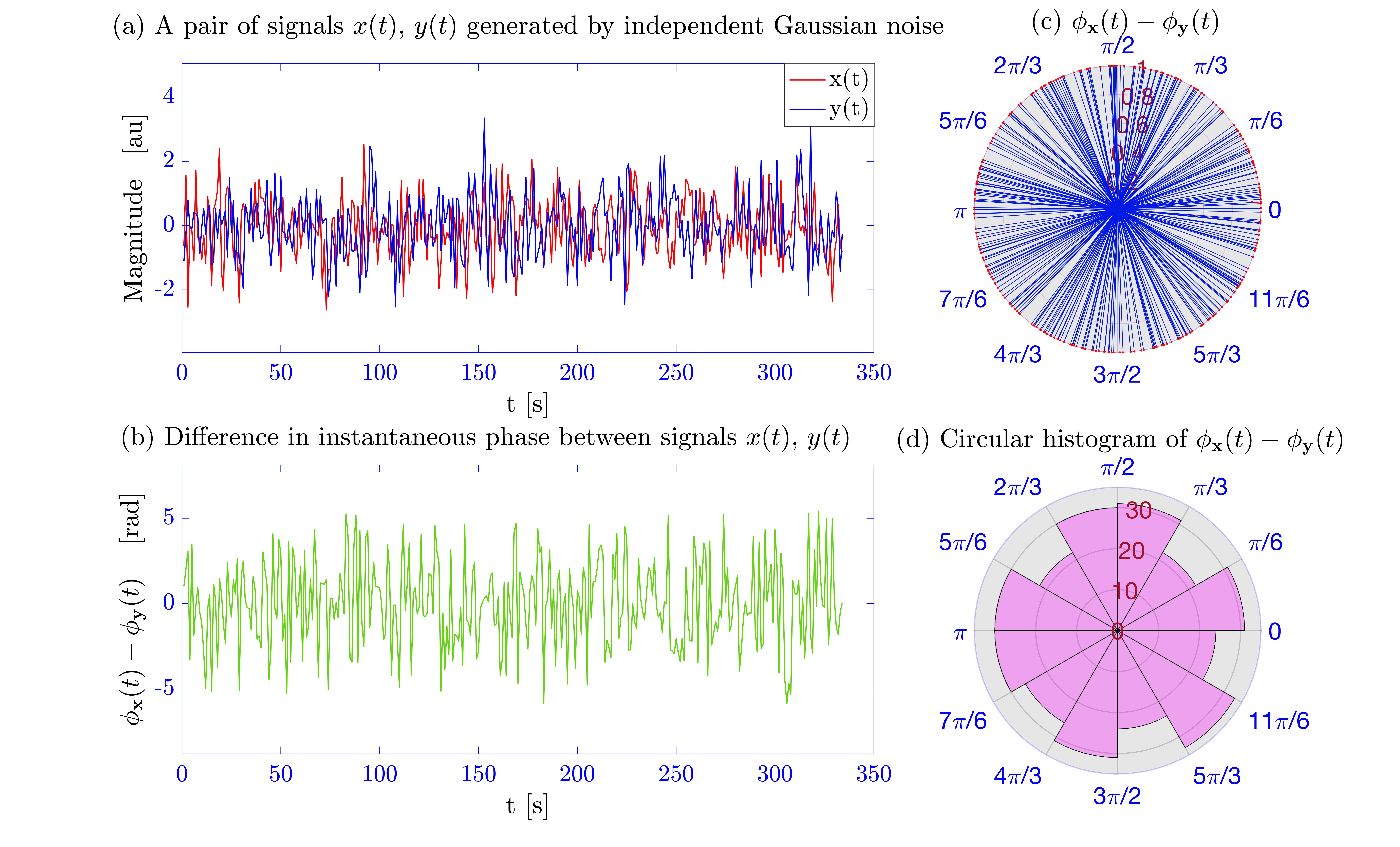}
\end{center}
\caption{A single realization of Simulation  1. (a) A pair of signals $x(t)$ and $y(t)$  generated from an independent Gaussian process. (b) The difference in the estimated phase between the signals at each time point. (c) The circular distribution of the phase difference time course in a polar coordinate system. (d) Same results in histogram form.}\label{fig: figBivariateNullDistribution}
\end{figure}

 Fig. \ref{fig: figBivariateNullDistribution} shows a single realization of Simulation 1 for illustration purposes.  Panel (a) shows a randomly generated pair of time series, and (b) the extracted instantaneous phases between the two time series at each time point. Since this is null data, the phase difference should vary uniformly in the interval $[0, 2\pi]$ as illustrated in Panels (c) and (d).
 
\begin{figure}[H]
\begin{center}
\includegraphics[width=\textwidth,trim={4cm 4.2cm 3.5cm 3.5cm},clip]{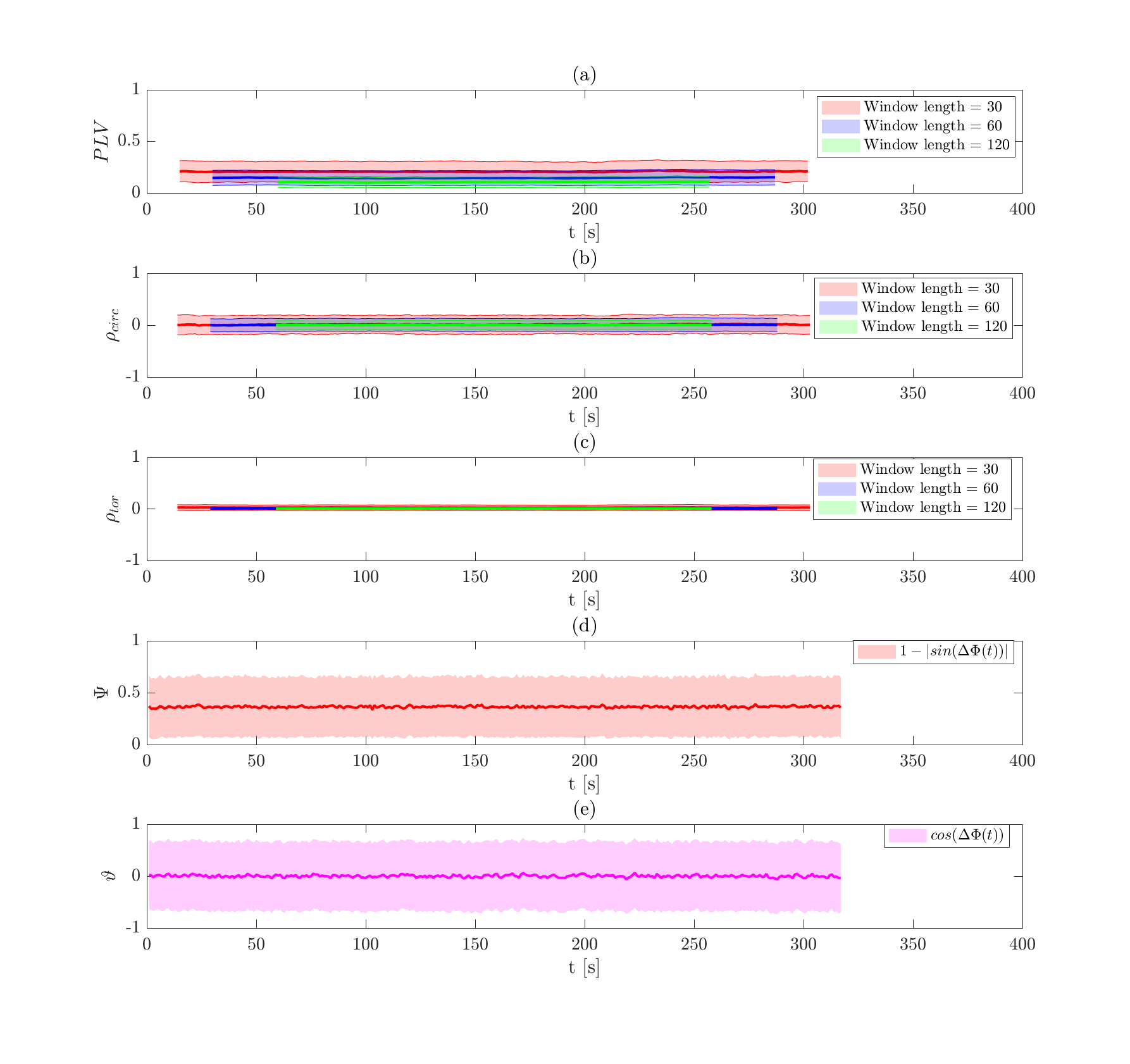}
\end{center}
\caption{Results of Simulation 1 without band-pass filtering.  The bold line indicates the estimated value, while the shaded area represents the 95\% confidence interval.  Results are shown for: (a) PLV using a sliding window; (b) circular-circular correlation using a sliding window; (c) toroidal correlation using a sliding window; (d) phase coherence; and (e) CRP. The sliding window techniques are evaluated at three different window lengths. }\label{fig: figbivarCOmparNullNofilter}
\end{figure}

Figure \ref{fig: figbivarCOmparNullNofilter}  shows a summary of the results for $1000$ surrogate data sets with the analysis performed on the data prior to band-pass filtering.  The mean and 95\% confidence interval for each measure are shown at each time point.  

Results for the WPS measures (Panels (a) - (c)) are shown for each window length (30, 60, and 120 time points). The results illustrate that all measures of PS are roughly constant across time. 

Note that measures such as PLV and phase coherence take values between $0$ and $1$. The mean value using phase coherence is roughly $0.35$ (Panel (d)), which is consistent with the results obtained using PLV for a window size of $60$ (Panel (a)). As the window size decrease, the value of PLV tends to be lower.  

In contrast, circular-circular correlation, toroidal circular correlation, and CRP all take values between $-1$ and $1$.  The mean of the CRP is $0$ at each time point (Panel (e)), which is consistent with the results of the WPS obtained using circular-circular and toroidal-circular correlation (Panels (b) and (c)).

It is important to note that phase coherence and CRP preserve the temporal resolution of the phase difference as they are not estimated using a sliding window. However, this appears to come at the cost of increased variability as indicated by the relatively wider $95\%$ confidence intervals. The effect of the chosen window length on various WPS measures shows that as the window size increases, the estimates converge towards their true values (i.e., $0$ for circular-circular correlation and toroidal circular correlation). 

\begin{figure}[H]
\begin{center}
\includegraphics[width=\textwidth,trim={3.1cm 4cm 4.6cm 3cm},clip]{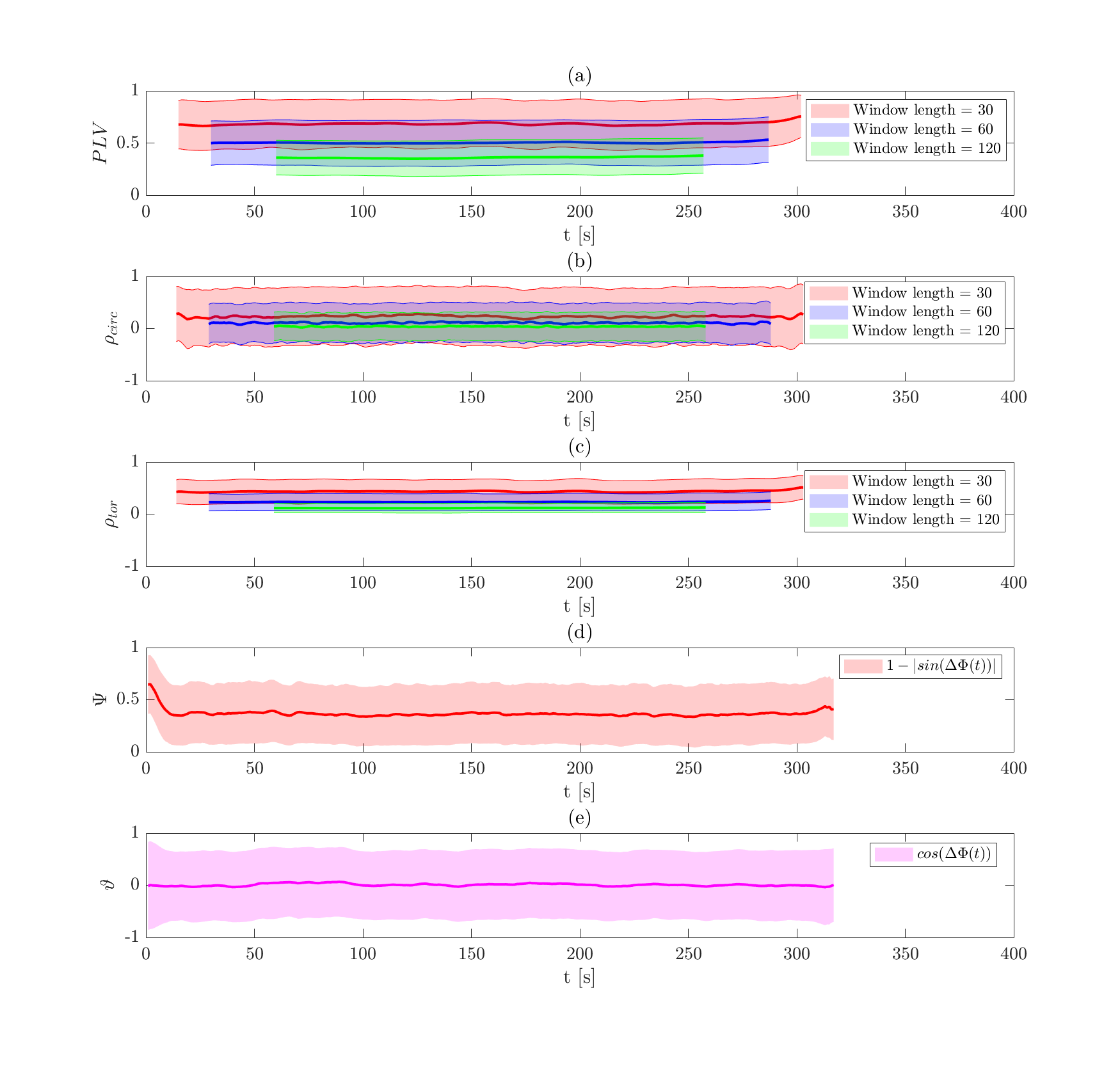}
\end{center}
\caption{Results of Simulation 1 with band-pass filtering.  The bold line indicates the estimated value, while the shaded area represents the 95\% confidence interval. Results are shown for: (a) PLV using a sliding window; (b) circular-circular correlation bsing a sliding window; (c) toroidal correlation using a sliding window; (d) phase coherence; and (e) CRP. The sliding window techniques are evaluated at three different window lengths.} \label{fig: fig5}
\end{figure}

Fig. \ref{fig: fig5} shows the comparison between various measures of PS used on the surrogate data after band-pass filtering.  The results indicate that synchronization measures remain roughly constant across time. However, the WPS measures (Panels (a) - (c)) show a noticeable difference compared to the results without band-pass filtering. Significantly, the results of the WPS measures show inflated values, indicating a higher degree of phase synchronization than would be expected in a null setting. PLV, circular-circular correlation, and toroidal correlation take values around $0.84$, $0.53$, and $0.62$, respectively.   These is driven by the fact that both signals have a center frequency of $0.05 \, Hz$ after applying band-pass filtering, leading to a situation where the signals are constrained to remain phase locked throughout the time course.  The IPS measures are not similarly affected. The mean value using phase coherence is again roughly $0.35$ (Panel (d)), and the mean of the CRP is $0$ at each time point (Panel (e)). Both values are roughly equivalent to those seen before band-pass filtering.

\subsection{Simulation 2}

\begin{figure}[H]
\begin{center}
\includegraphics[width=\textwidth,trim={2cm 3cm 2cm 6cm},clip]{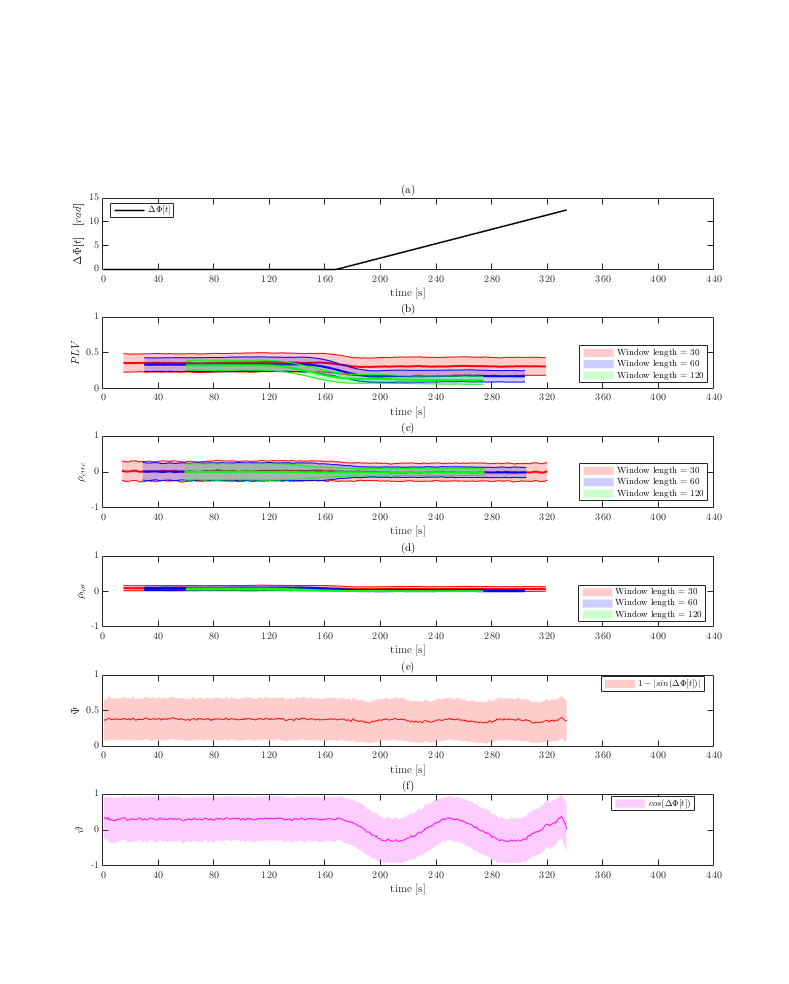}
\end{center}
\caption{Results of Simulation 2 without band-pass filtering.  (a) The ground truth phase shift between the two signals as a function of time. 
 Results are shown for: (b) PLV using a sliding window; (c) circular-circular correlation using a sliding window; (d) toroidal correlation using a sliding window; (e) phase coherence; and (f) CRP. The sliding window techniques are evaluated at three different window lengths.}\label{fig: BivariateCorrRamp_nobandpass}
\end{figure}

\begin{figure}[H]
\begin{center}
\includegraphics[width=\textwidth,trim={4cm 5cm 4cm 4cm},clip]{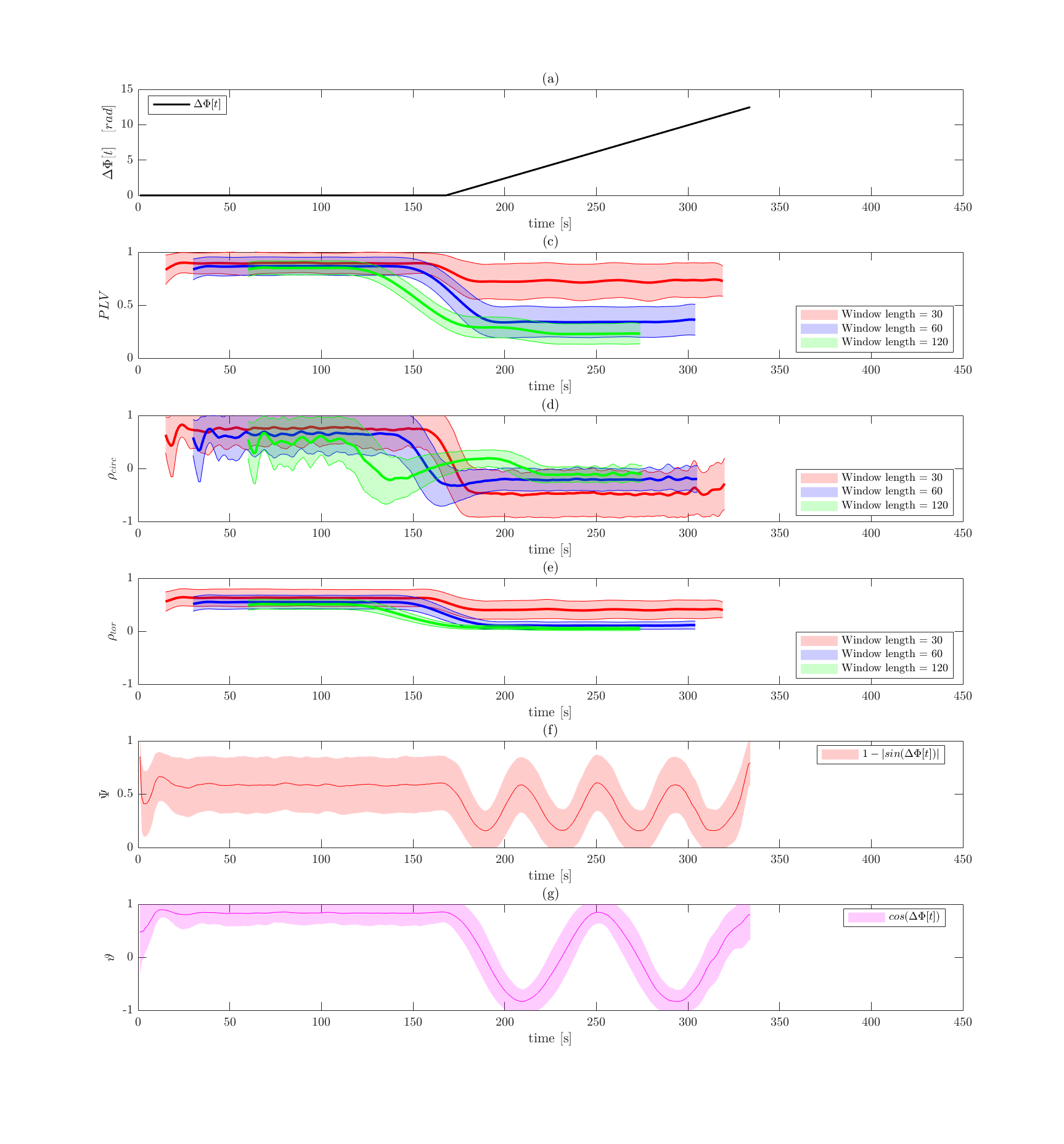}
\end{center}
\caption{Results of Simulation 2 with band-pass filtering.  (a) The ground truth phase shift between the two signals as a function of time. 
Results are shown for: (b) PLV using a sliding window; (c) circular-circular correlation using a sliding window; (d) toroidal correlation using a sliding window; (e) phase coherence; and (f) CRP. The sliding window techniques are evaluated at three different window lengths.}\label{fig: BivariateCorrRamp}
\end{figure}

Figures \ref{fig: BivariateCorrRamp_nobandpass} and \ref{fig: BivariateCorrRamp} illustrate the results of Simulation 2 performed on the data before and after band-pass filtering, respectively. Recall that in this simulation the two signals are designed to have the same phase up to time $t=170$, after which a phase shift is introduced that varies linearly from $0$ to $4\pi$ (see Figures \ref{fig: BivariateCorrRamp_nobandpass}a). Thus, the signals should gradually move in and out of phase during the second half of the time course. Here the signals will be in-phase when the phase difference is $2\pi$ and $4\pi$, and in anti-phase when the difference is $\pi$ and $3\pi$.

The costs of not band-pass filtering the data are apparent in Figure \ref{fig: BivariateCorrRamp_nobandpass}, as all five of the methods return results consistent with those seen in the null setting. None of the methods does a good job of either detecting the fact that the signals are in phase in the first half of the time course, or that they gradually move in and out of phase in the second half. This can be explained by the fact that the signal is contaminated with noise from all frequencies, which in turn corrupts the estimated instantaneous phase. 

Contrast this with the results after band-pass filtering shown in Figure \ref{fig: BivariateCorrRamp}. Here all of the measures of PS correctly predict a value close to $1$ in the first half of the signal, indicating that all methods are picking up on the fact that the signals are in phase.  In Panels (b)-(d), which represents the WPS measures, the phase shift occurring after $t=170$ leads to a decrease in phase synchronization from this time point on.  The toroidal correlation appears to perform best, showing more sensitivity in detecting the episodes of phase synchronization compared to PLV and circular-circular correlation. It can also be observed that the circular-circular correlation is more susceptible and sensitive to the noise than the other measures (see the increased wiggles in the estimates values). Interestingly, the PLV results appear to be more sensitive to the window length used than the other two metrics. However, it is important to note that none of the WPS methods are able to detect that the signals are in phase when the phase shift equals $2\pi$ and $4\pi$. 

In contrast, both the phase coherence and CRP better captures the PS variation than the WPS measures. This is partly due to the fact that using a sliding window deteriorates the resolution of the PS depending on the window size. In particular, note how well CRP detects that the signal is in phase at points when the phase shift equals $2\pi$ and $4\pi$. This is in contrast to phase coherence that erroneously assumes that signals are also in phase when the shifts are equal to $\pi$ and $3\pi$. The latter is due to the fact that phase coherence cannot differentiate between when the signals are in phase from when they are in anti-phase.

\subsection{Simulation 3}\label{sec: bivarsigmoid}

\begin{figure}[H]
\begin{center}

\includegraphics[width=\textwidth,trim={4.5cm 5cm 4.2cm 5cm},clip]{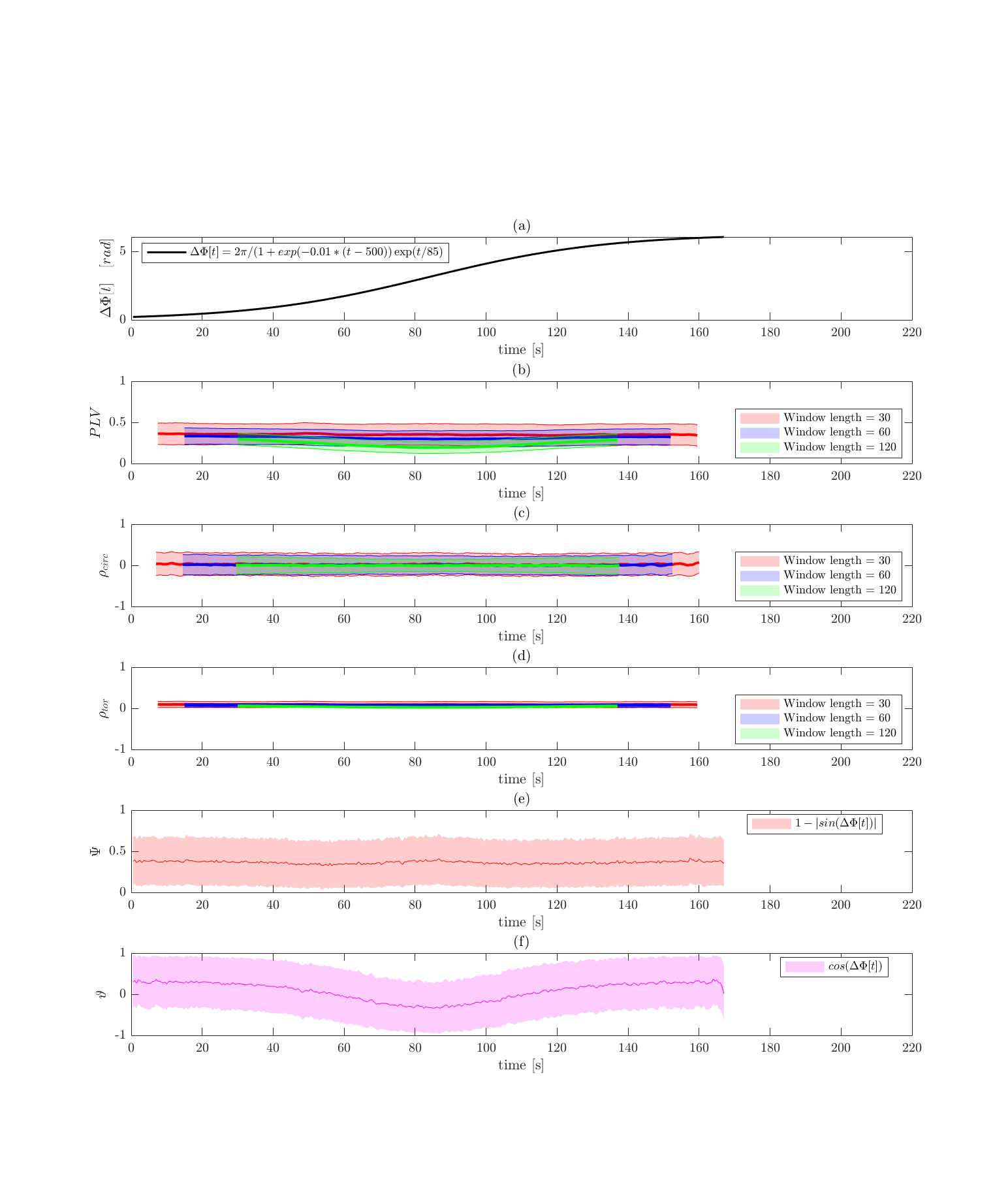}
\end{center}
\caption{Results of Simulation 3 without band-pass filtering.  (a) The ground truth phase shift between the two signals as a function of time. Results are shown for: (b) PLV using a sliding window; (c) circular-circular correlation using a sliding window; (d) toroidal correlation using a sliding window; (e) phase coherence; and (f) CRP. The sliding window techniques are evaluated at three different window lengths.} \label{fig: BivariateSigmoid_NoBPFilter}
\end{figure}

\begin{figure}[H]
\begin{center}

\includegraphics[width=\textwidth,trim={3.5cm 6cm 4cm 9.3cm},clip]{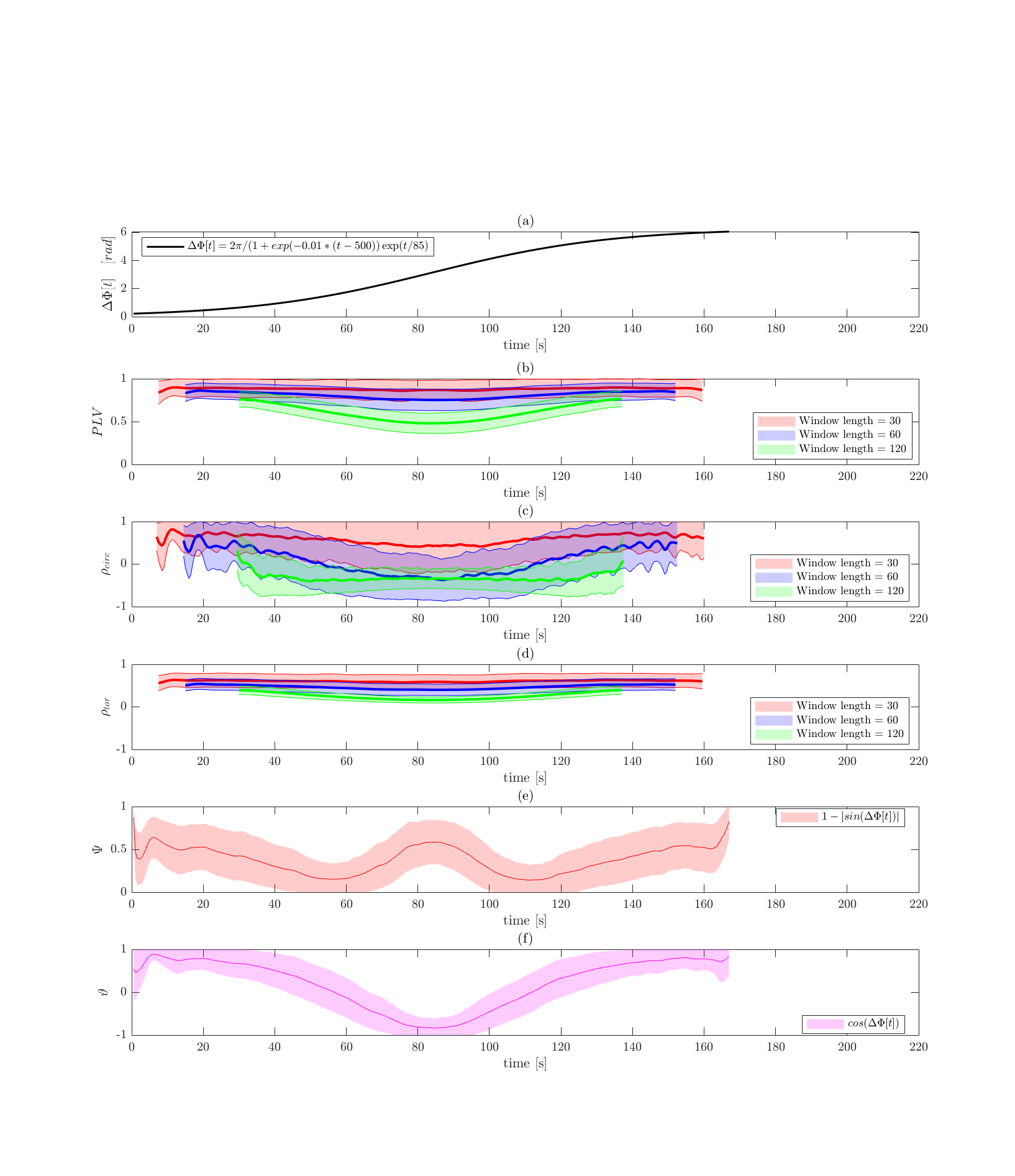}
\end{center}
\caption{Results of Simulation 3 with band-pass filtering.  (a) The ground truth phase shift between the two signals as a function of time.  Results are shown for: (b) PLV using a sliding window; (c) circular-circular correlation bsing a sliding window; (d) toroidal correlation using a sliding window; (e) phase coherence; and (f) CRP. The sliding window techniques are evaluated at three different window lengths.} \label{fig: BivariateSigmoid}
\end{figure}

Figures \ref{fig: BivariateSigmoid_NoBPFilter} - \ref{fig: BivariateSigmoid} illustrate the results of Simulation 3 performed on data before and after band-pass filtering. As illustrated in Figures \ref{fig: BivariateSigmoid_NoBPFilter}a, the two signals are designed to initially be in phase, after which they gradually go out of phase. At time $t=170$ when the phase difference is $2\pi$, the two signals will be in anti-phase, before returning to being in phase at the end of the time course.

The costs of not band-pass filtering the data are again apparent in Figure \ref{fig: BivariateSigmoid_NoBPFilter}, as all of the methods show results consistent with the null setting. This is in contrast to the results obtained after band-pass filtering shown in Figure \ref{fig: BivariateSigmoid}.  Here all measures of phase synchronization pick up on the fact that the signals start out in phase, gradually goes out of phase (culminating at time $t=170$), before gradually return to being in phase.  

In Panels (b)-(d), which represents the WPS measures, we see that using a longer window length tends to capture phase dynamics better than using a smaller window length.  Again, the toroidal correlation performs best, showing increased sensitivity in detecting the episodes of phase synchronization compared to circular-circular correlation and PLV. Increased window lengths provide better results.

Both phase coherence and CRP capture the manner in which phase synchonization varies more clearly than the WPS measures. In particular, CRP provides the most reliable measures in this simulation and clearly detects both when the signals are in and out of phase. In comparison, phase coherence cannot separate when the signals are in phase and anti-phase, illustrating one of the shortcomings of the approach.

\subsection{Analysis of Kirby Data}

\begin{figure}[H]
\begin{center}\includegraphics[width=0.5\textwidth,trim={0cm 1cm 0cm 0.1cm},clip]{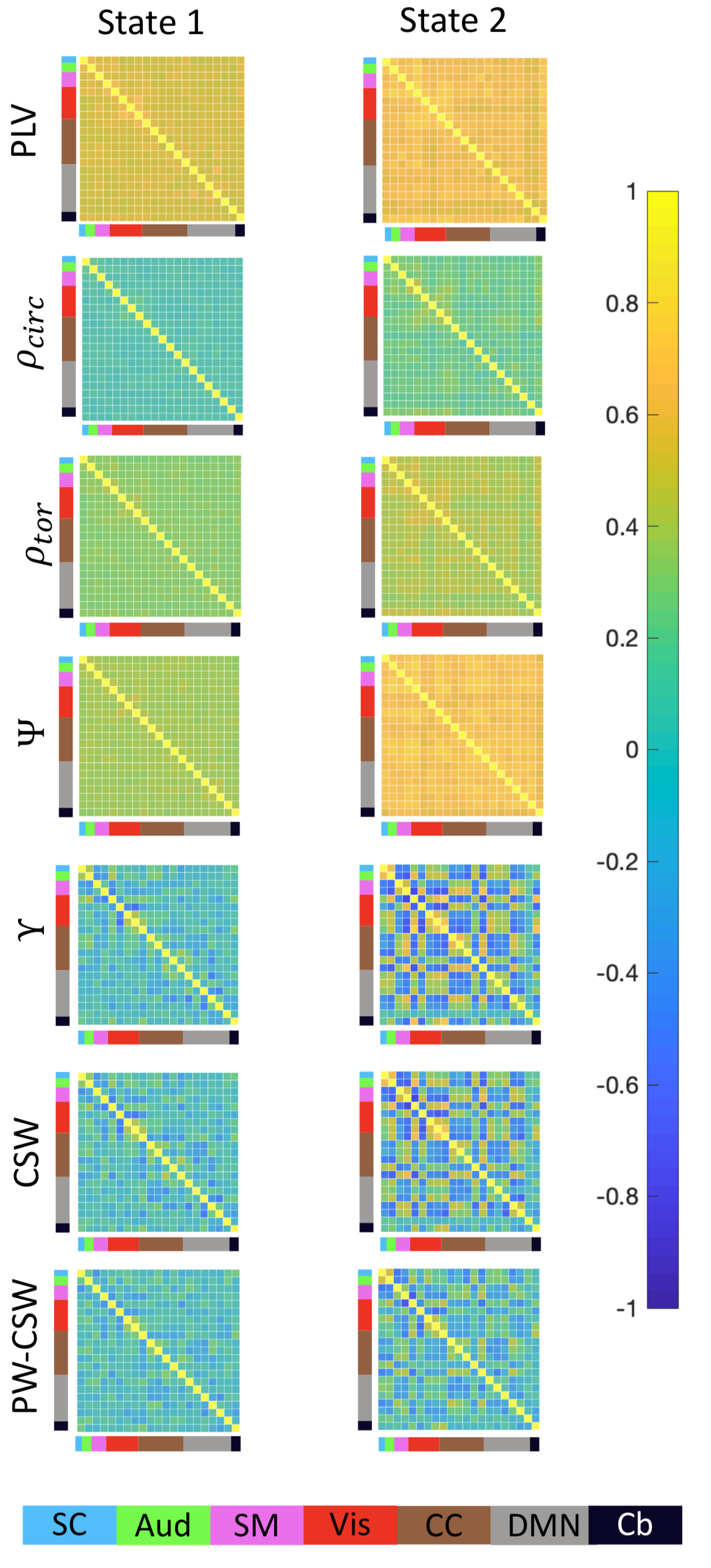}
\end{center}
\caption{Analysis of the Kirby21 Data.  After applying each method the time-varying connectivity measures were clustered into 2 reoccurring brain states.  Results are shown (top to bottom) for: PLV using a sliding window; circular-circular correlation using a sliding window; toroidal correlation using a sliding window; phase coherence; cosine of the relative phase; correlation-based sliding window; and prewhitened correlation-based sliding window. The sliding window techniques are evaluated with window length 30 time points.} \label{fig: Kirby21_6State}
\end{figure}

After applying each method to the rs-fMRI data, two brain states were extracted using k-means clustering. Figure \ref{fig: Kirby21_6State} contrasts the estimated brain states obtained using the different methods for assessing phase synchonization, as well as with correlation-based sliding window analysis (both with and without pre-whitening).  Brain states are organized into seven functional groups: visual (Vis); auditory (Aud); somatomotor (SM); default mode (DMN); cognitive-control (CC); sub-cortical (SC); and cerebellar (Cb) networks.

Beginning with the WPS methods, PLV (top row), circular-circular correlation (second row) and toroidal correlation (third row) show roughly similar results with regards to the relationship between functional groups in each brain state. However, the PLV derived brain states in general take higher values than those obtained using toroidal correlations, which in turn take higher values than those obtained using circular-circular correlations. These results are largely consistent with those seen in the simulation studies, and the fact that toroidal and circular-circular correlations take a wider range of values (compared to PLV which is constrained between $0$ and $1$).

Turning to the IPS methods, the brain states obtained using phase coherence (fourth row) tends to provide higher values in general compared to CRP (fifth row). This is not necessarily surprising as the range of potential values are different (phase coherence takes values between $0$ and $1$, while CRP takes values between $-1$ and $1$). In addition, as seen in the simulation studies phase coherence has problems differentiating between when signals are in phase versus when they are in anti-phase. Together, this provide higher values in the estimated brain states. For example, State 2 shows a hyper-connected where all PS measures are close to $1$.  

The final two rows of the figure show brain states estimated using sliding window correlations. Interestingly, the results obtained using sliding windows without any prewhitening (sixth row) is very similar to those obtained using the CRP. For both, State 2 was characterized by stronger correlations (both positive and negative) relative to State 1. Moderate to strong negative correlations between sensory systems (auditory, somatomotor, and visual) components were present in State 2 but were reduced in State 1. This similarity between methods indicates that CRP may be finding similar brains states as CSW, but using more high-resolution data as it does not use a predefined window. These findings are largely consistent with those of \cite{pedersen2018relationship} who found that IPS and CSW conveyed comparable information of time-resolved fMRI connectivity, though IPS provided finer temporal resolution.  Finally, the results obtained using sliding windows without prewhitening (seventh row) show lower estimated values than the results without prewhitening, which is consistent with results found in \cite{honari2019investigating}.

\section{Discussion}

There is growing interest in measuring time-varying functional connectivity between time courses from different brain regions using rs-fMRI data. One such approach is to measure their phase synchronization across time. In this paper, we evaluate a number of methods for measuring PS and contrast them with one another.  In discussing methods, we differentiate between two classes of methods: windowed phase synchronization and instantaneous phase synchronization.

WPS methods combine a static PS measure between two different signals with a sliding window to obtain a time-varying measure of PS. In principal, any metric that allows one to calculate an omnibus measure of PS can be used within this framework.  Since phase information is circular data, the use of circular-circular correlation and toroidal circular correlation were deemed natural candidate methods to use as a measure of PS.  To the best of our knowledge, neither approach has previously been used to study PS in fMRI.  The PLV in WPS method, has in contrast previously been used to assess episodes of elevated gastric-BOLD synchronization \citep{rebollo2018stomach}. 

IPS methods directly use the phase difference time series obtained from applying the Hilbert Transform, allowing one to compute an instantaneous measure of PS. This has the benefit of providing a higher temporal resolution, as there is no need to choose an arbitrary window size as for the WPS methods. However, there remains a related somewhat arbitrary choice of filter bandwidth to narrow-band the signals prior to analysis. Here we focused on two measures of IPS, phase coherence, which has already found wide usage in the field, and CRP, a newly developed method.

The three simulations illustrate several important points regarding the performance of these methods.  Simulation 1 shows that the WPS methods are highly affected by band-pass filtering. To illustrate, Fig. \ref{fig: fig5} shows that these methods tend to provide estimates that indicate that signals are consistently in phase, even when the phases are designed to randomly vary.   These results hold because the two signals being compared have a center frequency of $0.05 \, Hz$ after applying band-pass filter with cutoff frequencies $[0.03, 0.07] \, Hz$.  This leads to a situation where the signals are constrained to remain relatively phase locked and thus have constant PS throughout the time course.  Importantly, the IPS results appear to perform similarly on the data both before and after band-pass filtering (see Figs \ref{fig: figbivarCOmparNullNofilter} and \ref{fig: fig5}), and thus appear to be less sensitive to filtering in the null setting.

While at first glance, the results of Simulation 1 appear to indicate that band-pass filtering is not beneficial, and may in fact be detrimental, Simulations 2 and 3 put this notion to rest. Here, the results performed on the non band-pass filtered data indicate that none of the methods are able to pick up changes in real PS present in the data, and instead appear to erroneously indicate that the data behave in manner consistent with null data. This is largely corrected after band-pass filtering the data (see Figs. \ref{fig: BivariateCorrRamp} and \ref{fig: BivariateSigmoid}). This result holds both for WPS and IPS methods, and indicates that band-pass filtering is a necessary step in the analysis of PS.  

This result corresponds to theoretical findings (Bedrosian's theorem) that suggest using band-pass filters in the study of PS is critical for the signal to have physically meaningful demodulation into its envelope and instantaneous phase components. 
 
However, it is important to note that band-pass filtering comes at the cost of introducing further autocorrelation into the phase of the signal.  In addition, band-pass filtering increases the risk of spurious detection of phase synchronization \citep{rosenblum2000detection}. While a band-pass filter denoises the signal, it can also lead to an increase in the degree of synchronization by narrowing the band width; see Fig. \ref{fig: fig5}.

The results of Simulations 2 and 3 together show that all methods to a certain extent were able to detect changes in PS. Focusing on the WPS measures, toroidal correlation performed best, showing increased sensitivity to detecting episodes of PS compared to PLV and circular-circular correlation. Circular-circular correlation was the most susceptible and sensitive to noise. A previous study comparing PLV and circular-circular correlation \citep{pauen2013circular} suggested that circular-circular correlation is appropriate for estimating the phase coupling reliably and not restricted to bivariate analyses.  It also indicated that using it as a measure of phase coupling could show slightly lower estimates than its counterpart. This result is consistent with what we found in our simulations.  

When assessing WPS measures, we also investigated a variety of window lengths.  The simulations indicated that shorter windows yielded a higher estimate of phase synchronization and increased risk of detecting spurious phase synchronization. However, longer windows made it harder to detect subtle changes. In general, longer window lengths tend to provide more accurate estimates of PS as they lead to a decrease in the variation of the estimates.  

IPS measures consistently outperformed WPS measures in the simulations, and were able to better pick up changes in PS across time. While phase coherence offers more accurate and sensitive results than the WPS methods, it still discards information about the direction of the relationship.  In contrast, CRP was not only able to detect phase synchronization but also preserved the directional information contained in the relative phase difference of the signals. However, one should note that the variation present in the IPS methods appears larger than WPS methods as evidenced by the wider confidence bounds.

It is interesting to consider the range of values each method returns. PLV and phase coherence take values between $0$ and $1$.  In contrast, circular-circular correlation, toroidal circular correlation, and CRP all take values between $-1$ and $1$.  This has to be taken into consideration when interpreting the results of each method. For example, in Simulation 1 where we analyzed null data, the latter methods returned values that lay symmetrically around $0$. This makes it easier to interpret null results compared to PLV and phase coherence whose null values were around $0.35$. Understanding what null values look like is a critical component towards understanding the performance of a method, as it is otherwise difficult to differentiate signal from noise.

Application to real data showed results that were consistent with the simulations. The WPS methods showed roughly equivalent results with respect to the relationship between functional groups in each estimated brain state.  As described above, the shorter range of values for PLV and phase coherence made it more difficult to pick up subtle differences between brain states, and they both returned a hyper-connected brain state where all PS measures are close to 1. Interestingly, the brain states estimated using CRP closely resembled those estimated using sliding window correlations. Thus, it appears that CRP is finding similar brains states but using more high-resolution data as it does not require the use of a sliding window.

In summary, we recommend the use of CRP as a measure of PS as it is able to separate when the signals are in phase from when they are in anti-phase. In addition, it returns a range of values similar to correlation, which makes it possible to interpret results similarly. Of all the methods tested, it showed the greatest concurrence with CSW, with the benefit of not having to predefine a window length.

\section*{Acknowledgments}

The work presented in this paper was supported in part by NIH grants R01 EB016061 and R01 EB026549 from the National Institute of Biomedical Imaging and Bioengineering and R21 NS104644 from the National Institute of Neurological Disorders and Stroke.

\newpage

\end{document}